\documentclass[sigconf,natbib=true,anonymous=false]{acmart}

\usepackage{xspace}
\usepackage{mathtools}
\usepackage{algorithm}
\usepackage{algpseudocode}
\usepackage{amsmath}
\usepackage{amsfonts}
\usepackage{tikz}
\usepackage[show]{chato-notes}
\usepackage{enumitem}
\usepackage{multirow}
\usepackage{graphicx}
\usetikzlibrary{positioning, arrows.meta, shapes.multipart}
\usepackage{booktabs}
\usepackage{makecell}
\usepackage{fontawesome5}

\AtBeginDocument{%
  }

\setcopyright{acmlicensed}
\copyrightyear{2018}
\acmYear{2018}
\acmDOI{XXXXXXX.XXXXXXX}
\acmConference[SIGIR 26]{}{June 03--05,
  2018}{Woodstock, NY}

\newcommand{\splade}{\textsc{Splade}\xspace}
\newcommand{\sparton}{\textsc{Sparton}\xspace}
\newcommand{\logp}{log1p\xspace}

\newcommand{\spladet}{\textsc{Splade}-V3\xspace}

\begin{document}

\title{Sparton: Fast and Memory-Efficient Triton Kernel for Learned Sparse Retrieval}

\author{Thong Nguyen}
\orcid{1234-5678-9012}
\affiliation{%
  \institution{University of Amsterdam}
  \city{Amsterdam}
  \country{Netherlands}
}
\email{t.nguyen2@uva.nl}

\author{Cosimo Rulli}
\affiliation{%
  \institution{ISTI-CNR}
  \city{Pisa}
  \country{Italy}}
\email{cosimo.rulli@isti.cnr.it}

\author{Franco Maria Nardini}
\affiliation{%
  \institution{ISTI-CNR}
  \city{Pisa}
  \country{Italy}
}
\email{francomaria.nardini@isti.cnr.it}

\author{Rossano Venturini}
\affiliation{%
 \institution{University of Pisa}
 \city{Pisa}
 \country{Italy}}
 \email{rossano.venturini@unipi.it}

\author{Andrew Yates}
\affiliation{%
  \institution{Johns Hopkins University, HLTCOE}
  \city{Baltimore}
  \state{MD}
  \country{USA}}
\email{andrew.yates@jhu.edu}

\setcopyright{none}
\settopmatter{printacmref=false, printccs=false, printfolios=true}
\renewcommand\footnotetextcopyrightpermission[1]{}



\begin{abstract}

State-of-the-art Learned Sparse Retrieval (LSR) models, such as \splade, typically employ a Language Modeling (LM) head to project latent hidden states into a lexically-anchored logit matrix. This intermediate matrix is subsequently transformed into a sparse lexical representation through element-wise operations (ReLU, \logp) and max-pooling over the sequence dimension.
Despite its effectiveness, the LM head creates a massive memory bottleneck due to the sheer size of the vocabulary ($\mathcal{V}$), which can range from 30,000 to over 250,000 tokens in recent models. Materializing this matrix creates a significant memory bottleneck, limiting model scaling. The resulting I/O overhead between operators further throttles throughput and runtime performance.  
In this paper, we propose \sparton, a fast---memory-efficient---Triton kernel tailored for the LM head in LSR models. \sparton utilizes a fused approach that integrates the tiled matrix multiplication, ReLU, \logp, and max-reduction into a single GPU kernel. By performing an early online reduction directly on raw logit tiles, \sparton avoids materializing the full logit matrix in memory. Our experiments demonstrate that the \sparton kernel, in isolation, achieves up to a $4.8\times$ speedup and an order-of-magnitude reduction in peak memory usage compared to PyTorch baselines. Integrated into \splade ($|~\mathcal{V}| \approx 30\text{k}$), \sparton enables a 33\% larger batch size and 14\% faster training with no effectiveness loss. On a multilingual backbone ($|\mathcal{V}| \approx 250\text{k}$), these gains jump to a $26\times$ larger batch size and $2.5\times$ faster training. 

\begin{center}
    \faGithub \,\, \url{https://github.com/thongnt99/sparton}
\end{center}

\end{abstract}

\maketitle


\section{Introduction}
\label{sec:intro}

\begin{figure}
 \centering
 \includegraphics[trim={0 0.73cm 0 0},clip, width=0.9\linewidth]{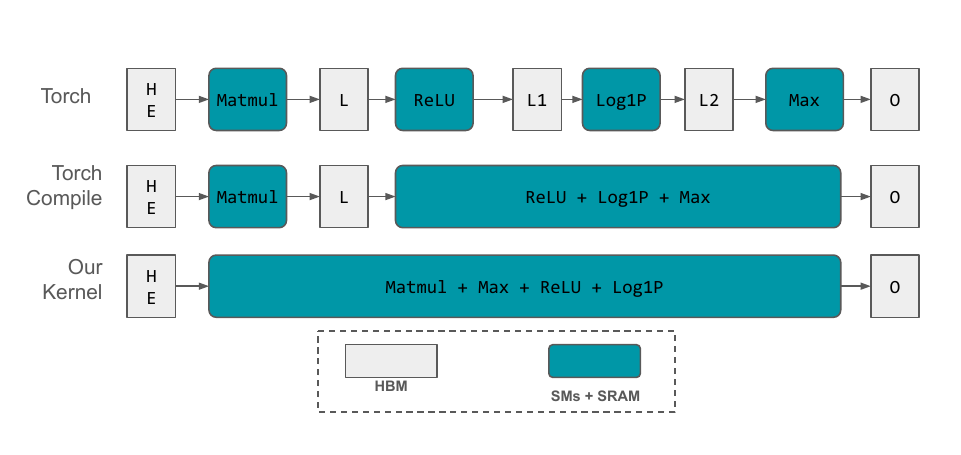}
 \caption{LM implementations in PyTorch and \sparton. Data in HBM (grey) is loaded into SRAM per block for parallel computation in Streaming Multiprocessors (green).}
 \vspace{-0.5cm}
 \label{fig:lsr_kernels}
\end{figure}

Learned Sparse Retrieval (LSR)~\cite{epic, lassance2024splade, lassance2022efficient-splade, geng2024towards, nguyen2025milco} has emerged as a powerful paradigm for producing sparse, high-dimensional representations aligned with the vocabulary of a pretrained language model. By capturing rich semantic relationships between terms, LSR bridges the gap between lexical and neural retrieval, rivaling the effectiveness of dense and multivector encoders. Moreover, thanks to tailored inverted indexes for learned sparse representations~\cite{bruch2024seismic, bruch2024pairing}, LSR embeddings can be retrieved as efficiently as dense embeddings served by advanced data structures such as HNSW~\cite{hnsw}.

The main gap between LSR methods and their dense or multivector counterparts lies in training efficiency. A key bottleneck is the Language Model (LM) head, a module that projects the hidden states produced by the transformer backbone into a sparse vector representation anchored in a lexical vocabulary. In state-of-the-art LSR models such as \splade~\cite{formal2021splade, lassance2024splade}, this component introduces substantial computational and memory overhead.

As a first step, considering a batch of $B$ input sequences of size $S$ on a token vocabulary $\mathcal{V}$, the LM head computes and materializes the logits matrix $L \in \mathbb{R}^{B \times S \times |\mathcal{V}|}$. Then, two element-wise operations---ReLU and \logp---followed by max-pooling over the sequence dimension, extract the salience scores of the most relevant terms, yielding a sparse output vector of $|\mathcal{V}|$ dimensions.

The bottleneck arises due to how this pipeline is implemented in standard frameworks such as PyTorch~\cite{paszke2019pytorch}. These operations execute sequentially: the logit matrix is computed block-wise in fast on-chip memory (SRAM) via an optimized matmul kernel, but must be written back to slower global memory (HBM) due to its large size. It is then repeatedly transferred between SRAM and HBM for each subsequent operation (ReLU, \logp, max-pooling).

Moving data back and forth between SRAM and HBM is a highly suboptimal approach that fails to account for the memory-bound nature of the problem. First, materializing the full logit matrix in HBM creates an unwanted memory peak; on a BERT-like architecture, considering $B=S=512$ and $|\mathcal{V}| \approx 30k$, the logit matrix requires $16$ GB to be stored. Observe that while modern GPUs can have up to 80 Gbytes of HBM, the SRAM is in the order of hundreds of kilobytes. This limits the maximum batch size, which is known to positively correlate with training speed and the effectiveness of the model~\cite{gao2021scaling}.
In terms of speed, repeatedly reading this large logit matrix from HBM for processing in SMs and writing intermediate results back introduces significant delays due to data movement. This limits the exploitation of massive arithmetic throughput enabled by the thousands of parallel cores available on A100 or H100~\cite{huerta2025dissecting, luo2024benchmarking}.

In this paper, we introduce \sparton, a fast---memory-efficient---kernel implemented in Triton~\cite{tillet2019triton} tailored for the LM head in LSR models. Specifically, \sparton fuses all operations performed by the LM head, i.e., matmul, max, ReLU, and \logp, on the logit matrix into a single Triton kernel (Figure \ref{fig:lsr_kernels}). During the forward pass, we perform on-chip, online max-reduction, fused with ReLU and \logp, so that we write only the max values and their indices to HBM. This eliminates the need to store the full intermediate logit matrix, dramatically reducing both memory usage and data movement overhead.
For the backward pass, we reuse the stored max values and their indices to compute gradients for the hidden states and vocabulary embeddings. By doing so, we require only the hidden states at the max indices to be loaded and differentiated, saving additional compute and memory bandwidth. 

Our experiments demonstrate that our \sparton kernel, in isolation, achieves up to a $4.8\times$ speedup and an order-of-magnitude reduction in peak memory usage compared to a PyTorch baseline. When integrated into the full \splade model ($|\mathcal{V}| \approx 32\text{k}$) , \sparton, despite replacing only the final component, enables training with a 33\% larger batch size (512 vs. 384), while simultaneously achieving 14\% faster training with no accuracy loss compared to a compiled PyTorch baseline. On a multilingual backbone (\texttt{xlm-roberta-base}, $|\mathcal{V}| \approx 250\text{k}$), we even see higher gains with a $26\times$ larger batch size and $2.5\times$ training speedup.

\section{LM Sparse Encoder}

Sparse Encoders encode queries and documents into sparse lexical-grounded representations; While several architectures exist, such as Binary and MLP encoders~\cite{nguyen2023unified}, the Language Model (LM) encoder is arguably the most effective and is therefore commonly adopted in state-of-the-art LSR models like \splade. The LM encoder comprises the pre-trained transformer backbone (e.g., \textsc{Bert}) and the LM head.  The backbone encodes raw text into a hidden state tensor $H \in \mathbb{R}^{B \times S \times D}$, where $B$ is the batch size, $S$ is the sequence length, and $D$ is the hidden dimension.  The LM sparse head consists of the standard language modeling head, which produces logits over a vocabulary, followed by several element-wise operators (ReLU, \logp) and a max pooling operator. The final output is a batch of vocabulary-sized lexical vectors whose sparsity (i.e., ratio of zeros) is induced by sparse regularizers during training. 

Formally, let $E \in \mathbb{R}^{|\mathcal{V}| \times D}$ denote the vocabulary embedding matrix, $b$ the bias vector, and $M \in \{0, 1\}^{B \times S}$ the attention mask. The sparse output embeddings  $\mathbf{Y} \in \mathbb{R}^{B \times |\mathcal{V}|}$ is computed as:
\begin{equation}
\mathbf{Y} = \max_{s} \left[ \log\left(1 + \text{ReLU}(H E^\top + b) \right) \odot M' \right]
\label{eq:mlm_head}
\end{equation}
where $M' \in \mathbb{R}^{B \times S \times |\mathcal{V}|}$ is the broadcasted over the vocabulary dimension, $\odot$ is the Hadamard product, and the max operation aggregates over the sequence dimension $S$.

\begin{algorithm}[t!]
\caption{Standard LM Head implementation.}
\begin{algorithmic}[1]
\Require $H \in \mathbb{R}^{B \times S \times D}$, $E \in \mathbb{R}^{|\mathcal{V}| \times D}$,  $b \in \mathbb{R}^{|\mathcal{V}|}$, $M \in \{0, 1\}^{B \times S}$
\State Read $H, E$ from HBM, compute $L = HE^{\mathsf{T}}$, write $L$ to HBM
\State Read $L, b$ from HBM, compute $L = L + b$, write $L$ to HBM
\State Read $L, M$ from HBM, compute $L = L \odot M$, write $L$ to HBM
\State Read $L$ from HBM, compute $L = ReLU(L)$, write $L$ to HBM
\State Read $L$ from HBM, compute $L = log(L+ 1)$, write $L$ to HBM
\State Read $L$ from HBM, compute $Y = \text{max}_{s} L$, write $Y$ to HBM
\end{algorithmic}
\label{algo:mlm_standard}
\end{algorithm}

In the standard Pytorch Eager mode, the LM head computation is executed sequentially as described in Algorithm \ref{algo:mlm_standard}.
As mentioned in Section~\ref{sec:intro}, the matrix $L$ can require up to $16$ GByte to be stored with common values of batch size and sequence length ($B=S=512$) on a BERT-like architecture.

Recent LLMs can process much longer sequences (e.g., 1024) and have larger vocabularies (e.g., 250k), which could lead to a massive memory bottleneck due to $L$. In every one of the next operators (steps 2 to 6), this logit matrix is repeatedly loaded and/or stored in the HBM, which entails a large number of memory accesses, resulting in significant runtime overhead.

Recent PyTorch versions ($\geq 2.0$) offer automatic compilation of Python code into a fused Triton~\cite{tillet2019triton} kernel to reduce data movement overhead.
Our investigation shows that this automatic compilation could fuse element-wise operators from step 2 to step 6. However, it does not fuse the matmul operation in step 1. For efficiency, PyTorch relies on third-party matrix-matrix multiplication black-box libraries (e.g., cuBLAS, rocBLAS)~\cite{paszke2019pytorch, markidis2018nvidia}, which do not allow customization. Therefore, the intermediate matrix $L$ is still materialized to the HBM, leaving the memory bottleneck unsolved. A similar memory bottleneck is also observed in the backward pass.

\section{\sparton: Efficient LM Head in Triton}
This section details \sparton's algorithms, which leverage the monotonicity of the involved operators and the sparse activation of the logits to reduce computation and alleviate the memory bottleneck in both the forward and backward passes of the LM head.

\smallskip
\noindent \textbf{Forward Pass}.
In the sparse representation formulation (Eq.~\ref{eq:mlm_head}), the pointwise mapping
$f(x)=\log(1+\mathrm{ReLU}(x))$ is monotonically non-decreasing. Therefore, for each $(b,v)$,
$\max_{s} f(\ell_{b,s,v}) = f(\max_{s} \ell_{b,s,v})$, where $\ell_{b,s,v}$ denotes the (masked, biased) logit.
This allows us to move the $\max$ reduction immediately after masking (step 3 in Alg.~\ref{algo:mlm_standard})
without changing the output. As a result, $\mathrm{ReLU}$ and \logp are applied to a $B\times |\mathcal{V}|$
tensor instead of a $B\times S\times |\mathcal{V}|$ tensor, reducing both arithmetic and data movement by a factor of $S$.
For example, with $B=S=512$, $|\mathcal{V}|=30522$, and half precision, activation I/O drops from $16$GB per pass  to approximately $31$MiB.

\begin{algorithm}[t!]
\caption{Forward Pass of \sparton.}
\begin{algorithmic}[1]
\Require $H \in \mathbb{R}^{B \times S \times D}$, $E \in \mathbb{R}^{|\mathcal{V}| \times D}$, $b \in \mathbb{R}^{|\mathcal{V}|}$,
$M \in \{0,1\}^{B \times S}$ 
\Ensure $Y \in \mathbb{R}^{B \times |\mathcal{V}|}$ and max indices $I \in \{0,\dots,S-1\}^{B\times |\mathcal{V}|}$
\For{each vocab tile $E_{\text{tile}} \in \mathbb{R}^{|\mathcal{V}_t| \times D}$ and $b_{\text{tile}} \in \mathbb{R}^{|\mathcal{V}_t|}$ of $b$}
    \State Read $H, E_{\text{tile}}, b_{\text{tile}}$ from HBM
    \State Compute
    $L_{\text{tile}} = H E_{\text{tile}}^{\mathsf{T}} + b_{\text{tile}}$ by a GEMM kernel
    \State Write $L_{\text{tile}}$ to HBM
    \State Read $L_{\text{tile}}, M$ from HBM
    \State Compute
    $Y_{\text{tile}}, I_{\text{tile}} = \log\!\left(1+\mathrm{ReLU}\!\left(\max\nolimits_s \left(L_{\text{tile}} \odot M\right)\right)\right)$
    \State Write $Y_{\text{tile}}, I_{\text{tile}}$ to HBM
\EndFor
\end{algorithmic}
\label{algo:sparton_forward}
\end{algorithm}

More importantly, the monotonicity reordering lets us accumulate the max on the raw logits, enabling a streaming
max-reduction over the sequence dimension where only the running maxima (and optionally argmax indices) are kept
on-chip; $\mathrm{ReLU}$ and \logp are applied once after reduction. Ideally, one would fuse GEMM, bias,
masking, and max reduction into a single kernel that performs this streaming reduction and writes only the reduced
$B\times |\mathcal{V}|$ outputs. We implemented this fully fused approach in Triton; while it is strictly more memory efficient as it avoids intermediate writes, the computational throughput of a custom Triton GEMM lags behind highly tuned GPU vendor libraries~\cite{markidis2018nvidia}. We therefore adopt a hybrid design: we compute tiled logits $L \in \mathbb{R}^{B\times S\times C}$ using
cuBLAS/rocBLAS over vocabulary tiles $E_{v_0:v_1}$, and immediately apply a Triton kernel to perform masked
$\max_s$ reduction (and optionally argmax) for each tile. We tile along the batch and vocabulary dimensions for block-wise kernel parallelism. Finally, we apply $\mathrm{ReLU}$ and \logp to
the reduced maxima on the same fused Triton kernel. This ensures the full $B\times S\times |\mathcal{V}|$ matrix is never
materialized while retaining optimized GEMM throughput. This approach is detailed in Algorithm \ref{algo:sparton_forward}.

\smallskip
\noindent \textbf{Backward Pass}.
Algorithm~\ref{algo:sparton_backward} computes gradients by exploiting the sparsity induced by the $\max_s$ operator. The forward pass stores only the reduced outputs per $(b,v)$—the max score $s_{\max}$ and the argmax
index $i_{\max}=\arg\max_s \ell_{b,s,v}$. During backward, the upstream gradient
$\delta=\nabla L[b,v]$ is multiplied by the derivative of $f(x)=\log(1+\mathrm{ReLU}(x))$, yielding
$g=\mathbf{1}[s_{\max}>0]\cdot \delta \cdot \exp(-s_{\max})$. The gradient then routes to a single hidden state
per $(b,v)$, namely $H[b,i_{\max}]$, and to the corresponding embedding vector $E[v]$, updating
$\nabla_E[v] \mathrel{+}= g\cdot H[b,i_{\max}]$ and $\nabla_H[b,i_{\max}] \mathrel{+}= g\cdot E[v]$ (with atomic
accumulation across parallel blocks). Notably, our backward pass is implemented as a \emph{single fused kernel} that
combines the chain rule through $\log(1+\mathrm{ReLU}(\cdot))$, argmax routing, and gradient accumulation, and
requires only $(s_{\max}, i_{\max})$ rather than the dense logits. The kernel is launched in parallel across $(b,v)$ blocks tiled along the batch and vocabulary dimensions. 

In contrast, both the standard eager PyTorch implementation and its auto-compiled variants (e.g., torch.compile)
treat GEMM as a black-box library call (cuBLAS/rocBLAS) and do not fuse the subsequent $\max_s$ reduction into the
GEMM output. Consequently, the dense logit tensor $L \in \mathbb{R}^{B\times S\times |\mathcal{V}|}$ must be materialized in HBM so that masking and the elementwise nonlinearity can be applied, and these
intermediates are typically retained for autograd. This incurs activation storage and bandwidth proportional to $\mathcal{O}(BS|\mathcal{V}|)$ even though $\max_s$ ultimately selects only one sequence position per $(b,v)$.
By saving only $(s_{\max}, i_{\max})$ and scattering gradients directly to the selected positions, our backward pass
reduces the saved forward state to $\mathcal{O}(B|\mathcal{V}|)$ and avoids excessive redundant activation storage, and its incurred I/O overhead.  

\begin{algorithm}[ht!]
\caption{Backward Pass of \sparton.}
\begin{algorithmic}[1]
\Require Upstream gradient $\nabla L \in \mathbb{R}^{B \times |\mathcal{V}|}$, Max scores and indices from forward pass, Inputs $H \in \mathbb{R}^{B \times S \times D}$, $E \in \mathbb{R}^{|\mathcal{V}| \times D}$
\Ensure Gradients $\nabla_H$, $\nabla_E$
\ForAll{batch block $b$ and vocab block $v$ in parallel}
    \State Read $\delta \gets \nabla L[b, v]$ from HBM
    \State Read $s_{\max}, i_{\max}$ from saved tensors on HBM
    \State Compute gradient through log1p-relu:
    \[
        g \gets 
        \begin{cases}
            \delta / \exp(s_{\max}), & \text{if } s_{\max} > 0 \\
            0, & \text{otherwise}
        \end{cases}
    \]
    \State Read embedding vector $e \gets E[v]$ from HBM.
    \State Read hidden state at max index: $h \gets H[b, i_{\max}]$ from HBM.
    \State Update sparse gradients: $\nabla_E[v] \mathrel{+}= g \cdot h$ ; $\nabla_H[b, i_{\max}] \mathrel{+}= g \cdot e$
\EndFor
\end{algorithmic}
\label{algo:sparton_backward}
\end{algorithm}

\section{Experiments}



\noindent \textbf{Experimental Setup.} We implement \sparton in Triton with optimal block tuning via Triton's \texttt{autotune} feature~\cite{tillet2019triton}. We compare it against PyTorch (Eager/Compiled) LM heads on NVIDIA A100/H100/200 GPUs. Metrics include latency ($ms$, via \texttt{perf\_counter} function of the \texttt{time} Python module) and peak memory (MiB, via \texttt{torch.cuda.}\texttt{max\_memory\_allocated()}), averaged over 20 iterations after 5 warm-up runs to exclude compilation overhead.

\vspace{2pt} 
\noindent \textbf{RQ1: What is the computational cost of the LM head in an LSR model?}
\begin{table}[t!]
    \centering
    \caption{Runtime ($ms$) and peak memory usage (MiB) of \spladet on a single H100 GPU ($B = 320, S = 512$).}
    \label{tab:splade_execution_comparison}
    \resizebox{\columnwidth}{!}{%
    \begin{tabular}{llrrrr}
        \toprule
        & & \multicolumn{2}{c}{\textbf{Eager Execution}} & \multicolumn{2}{c}{\textbf{Compiled Execution}} \\
        \cmidrule(lr){3-4} \cmidrule(lr){5-6}
        & \textbf{Component} & \textbf{Time (ms)} & \textbf{Mem (MiB)} & \textbf{Time (ms)} & \textbf{Mem (MiB)} \\
        \midrule
        \multirow{4}{*}{\rotatebox{90}{Fwd}} 
        & Backbone        & 84.4  & 2,893.7  & 99.7  & 3,083.6 \\
        & \quad + LM Head     & 162.1 & 28,885.1 & 122.1 & 10,126.0 \\
        & \quad + Tiled Head  & 140.3 & 5,310.7  & 123.1 & 3,101.3 \\
        & \quad + \sparton     & 113.7 & 2,955.4  & 129.0 & 3,146.0 \\
        \midrule
        \multirow{4}{*}{\rotatebox{90}{Fwd + Bwd}} 
        & Backbone        & 293.0 & 50,942.8 & 387.0 & 50,218.1 \\
        & \quad + LM Head     & 498.1 & 88,875.0 & 473.0 & 70,007.2 \\
        & \quad + Tiled Head  & 473.5 & 61,000.1 & 482.3 & 65,178.7 \\
        & \quad + \sparton     & 330.1 & 51,651.2 & 423.9 & 51,349.8 \\
        \bottomrule
    \end{tabular}
    }
\end{table}
As a first step in our evaluation, 
we break down the impact of the two main modules in a sparse encoder: (i) the transformer backbone and (ii) the Language Model (LM) head.
We use the state-of-the-art sparse encoder \spladet as a case study and measure its runtime and memory in both Pytorch eager mode and compiled mode (automatic compiling torch code to Triton kernel). Results reported in Table \ref{tab:splade_execution_comparison} show that the LM head induces a significant computational burden in both runtime and memory. 

In PyTorch's standard eager execution, the LM head overhead accounts for about the $48\%$ of the total latency, and it causes a drastic surge in memory usage, which jumps from 2,893.7 MiB (backbone only) to 28,885.1 MiB (full model), a $9.9$-fold increase. In the backward pass, the LM overhead accounts for $41\%$ in terms of latency and the $42\%$ in terms of memory.

Switching to \texttt{torch.compile} does not resolve these bottlenecks. While compilation provides some memory reduction for the forward pass (dropping to $\sim$10 GB), it fails to effectively alleviate the memory pressure during the backward pass, where usage remains critically high at $\sim$65 GB (15GB larger than the backbone). Furthermore, the runtime improvements are limited; the compiled full model (0.473s, fwd + bwd pass) is still significantly slower than the backbone alone. In fact, \texttt{torch.compile} slows down all backward passes, including on the backbone. These results indicate that automatic Pytorch's compliation optimizations are insufficient to mitigate the overhead of the LM head, highlighting the critical need for specialized techniques to optimize the LM head.

\vspace{2pt} 
\noindent \textbf{RQ2: Does tiling the Language Model head help?} 
We investigate the impact of tiling the logit computation alone (Algorithm~\ref{algo:sparton_forward}, line~1) to mitigate memory pressure.  As shown in Table \ref{tab:splade_execution_comparison}, our tiling proves to be highly effective for the forward pass in both execution modes. In the eager forward mode, it reduces total memory usage by a factor of $5.8\times$ (from 29GB to 5GB), with similar gains observed in the compiled version. The reason is that, with tiling, maximum memory usage is limited by the maximum tile size, which can be kept in the order of thousands. 
Yet, in the backward pass, using PyTorch’s autograd on the tiled forward computation fails to provide significant memory relief, with memory consumption remaining prohibitively high (over 61 GB in eager mode and 65 GB in compiled mode). Furthermore, it introduces a slight computational penalty in the compiled executions; the tiled head is slower than the vanilla LM head in both backward and forwards pass. These results confirm that neither tiling alone nor automatic PyTorch compilation can sufficiently resolve the training bottlenecks, motivating the need for a fully-fused custom kernel.

\vspace{2pt} 
\noindent \textbf{RQ3: How does our \sparton kernel perform compare to the tiled-only approach?}
\begin{figure}[t!]
    \centering
    \includegraphics[trim={0 0 0 0},clip, width=1.0\linewidth]{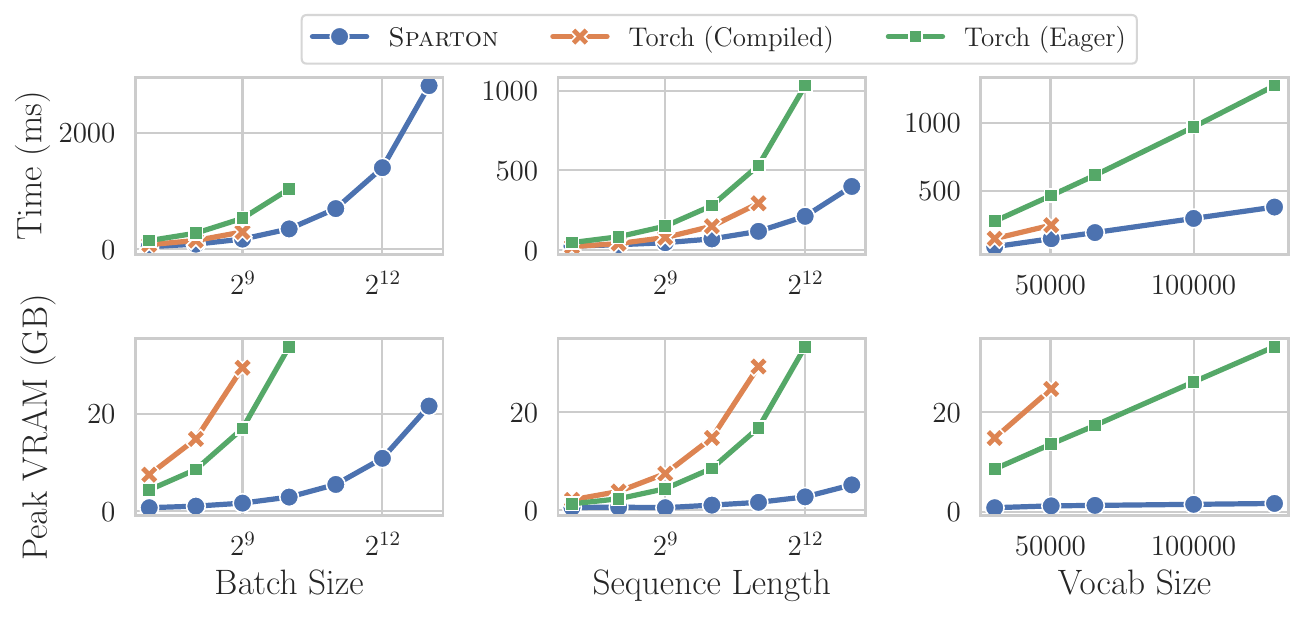}
    \caption{Scaling \sparton (without backbone) across three dimensions: Batch Size ($S=512, |\mathcal{V}|=30522$), Sequence Length ($B=128, |\mathcal{V}|=30522$), Vocabulary Size ($B=256, S=512$). }
    \label{fig:kernel_benchmarking}
    \vspace{-5pt}
\end{figure}
Beyond tiling, \sparton reorders the \texttt{max} operator ahead of the activation functions, reducing both I/O and computation on the subsequent operators.
In principle, this early reduction shrinks the stored activations by a factor of $S$, since only one value per sequence position needs to be retained for backpropagation.
In practice, however, PyTorch's autograd treats \texttt{matmul} as an opaque library call and still keeps its full output in HBM. 
\sparton tackles this limitation with a fully-fused backward kernel that stores only the max-reduced activations and computes gradients exclusively at the kept sequence positions, eliminating all redundant intermediate storage and I/O.
As shown in Table~\ref{tab:splade_execution_comparison}, these optimizations make \sparton the fastest and most memory-efficient implementation, nearly eliminating the LM head overhead relative to the backbone in both the forward and backward passes.

Figure~\ref{fig:kernel_benchmarking} reports latency (first row) and peak memory usage (second row) of the LM head (without the backbone) as a function of batch size ($B$), sequence length ($S$), and vocabulary size ($|\mathcal{V}|$), respectively, measured on an A100 40GB. \sparton outperforms the PyTorch baselines by a wide margin, with the gap widening as the input ($B$, $S$, or $|V|$) grows. 
While the baselines exhibit steep, linear (or worse) scaling, \sparton maintains a flat profile, making it increasingly superior for larger workloads.
At its best operating point, \sparton achieves up to a $4.8\times$ speedup and $12\times$ reduction in peak memory compared to compiled baseline.
We explicitly report some results in Table~\ref{tab:performance_comparison}, where we compare the latency and the peak memory usage as a function of the sequence length, while maintaining the batch size and the vocabulary size fixed to $128$ and $30522$, respectively. The table shows how \sparton is the only method that can handle input sequences of $8192$ values, while the other method fails. Both baselines hit a hard scalability wall: Tiled LM (compiled) encounters an out-of-memory (OOM) error at sequence length 4096, and Tiled LM (Eager) fails at 8192. In contrast, \sparton successfully scales to a sequence length of 8192 and beyond, utilizing only 5.13 GB of memory.

Overall, \sparton not only provides a significant speedup but also fundamentally solves the memory bottleneck, enabling much larger inputs where standard solutions fail.

\begin{table}[bt]
    \vspace{-5pt}
    \centering
    \caption{\sparton vs. Tiled LM with varying sequence lengths ($B=128, |V|=30522$) in the backward pass (on A100 GPU).}
    \label{tab:performance_comparison}
    \resizebox{\linewidth}{!}{%
    \begin{tabular}{lrrrrrr}
        \toprule
        \multirow{2}{*}{\bfseries\begin{tabular}{@{}c@{}}Seq\\Length\end{tabular}} & \multicolumn{2}{c}{\textbf{Tiled LM (Eager)}} & \multicolumn{2}{c}{\textbf{Tiled LM (Compiled)}} & \multicolumn{2}{c}{\textbf{\sparton (Ours)}} \\
        \cmidrule(lr){2-3} \cmidrule(lr){4-5} \cmidrule(lr){6-7}
         & Time (ms) & Mem (GB) & Time (ms) & Mem (GB) & Time (ms) & Mem (GB) \\
        \midrule
        1024 & 279.5 & 8.51 & 150.6 & 14.75 & \textbf{71.2} & \textbf{0.99} \\
        2048 & 531.1 & 16.82 & 294.2 & 29.37 & \textbf{118.6} & \textbf{1.55} \\
        4096 & 1031.0 & 33.44 & OOM & OOM & \textbf{212.7} & \textbf{2.68} \\
        8192 & OOM & OOM & OOM & OOM & \textbf{399.6} & \textbf{5.13} \\
        \bottomrule
    \end{tabular}%
    }
\end{table}

\subsection{End-to-End Training}

\begin{table}[b!]
\centering
\small
\setlength{\tabcolsep}{4pt}
\caption{Training efficiency and effectiveness (NDCG@10 on small-\textsc{Beir}) of \sparton in LSR training on NVIDIA H200. }
\label{tab:training_results}
\begin{tabular}{lrrrrr}
\toprule
Method & Batch & Steps & Time (h) & Mem (GB) & Avg. NDCG@10 \\
\midrule
\spladet & - & - &  - & - & 0.422 \\
\midrule
Compiled LM & 384 & 67528 & 14.24 & 125.78 & 0.421 \\
\sparton      & 384 & 67528 & 12.38 &  96.83 & 0.416 \\

\sparton     & 512 & 50646 & 12.24 & 128.63 & 0.427 \\

\bottomrule
\end{tabular}
\end{table}

In this section, we use our \sparton kernel to implement the LM head in an end-to-end real-world sparse encoder.

In doing so, we plug two different implementations of the LM head, namely the \sparton head and the PyTorch compiled head, downstream to the same backbone to compose our sparse encoder. 
We initialize the backbone using the same checkpoint that \splade-v3 was initialized from (\texttt{splade-cocondenser-selfdistil}, $|\mathcal{V}| \approx 30k$). We train our model on the Mistral-Splade data~\cite{doshi2024mistral}, using the InfoNCE loss~\cite{oord2019representationlearningcontrastivepredictive}, on a H200 GPU. To measure the kernel's correctness, we benchmark our encoders on a subset of the \textsc{Beir} datasets~\cite{thakur2beir} (small-\textsc{Beir}) collection consisting of the Arguana, FiQA, NFCorpus, SciDocs, and SciFact. We calculate and report the average NDCG@10 across the five datasets.

We report the results in Table \ref{tab:training_results}. When using the same batch size as the compiled LM, \sparton yields a similar retrieval effectiveness, confirming the correctness of our implementation. In terms of efficiency, \sparton is 14\% faster (12.24 hours vs. 14.24 hours) and uses nearly 30GB less memory than the compiled Pytorch version. Hence, we can scale the batch size up to 512 with \sparton. The increased batch-size leads to a slightly higher effectiveness for \sparton, matching the performance of \splade-V3 trained using a much more complex training recipe, involving knowledge distillation. 

When switching to a multilingual backbone (\texttt{xlm-roberta-base}, $|\mathcal{V}| \approx 250k $) while keeping the training setup the same,  we observe even a larger gain where \sparton enables $26 \times$ larger batch size (420 vs 16) and $2.5 \times$ faster training (67 vs. 25 hours) on an H100. 

\section{Conclusions and Future Work}
In this work, we present \sparton, an optimized Triton kernel that leverages tiled matrix multiplication, operator re-ordering, and sparse activation to alleviate memory bottlenecks and accelerate the LM head in LSR models. Our experiments demonstrate that \sparton provides a significant speedup while nearly eliminating the memory overhead of the LM head. This enables the processing of substantially larger batch sizes, sequence lengths, and vocabularies—scenarios where standard PyTorch implementations are either inefficient or encounter out-of-memory errors.  
Future work could extend \sparton to leverage hardware-specific features like Tensor Memory Accelerator (TMA) or lower-precision formats (e.g., FP8) to further exploit modern GPU capabilities.


\bibliographystyle{ACM-Reference-Format}
\bibliography{sample-base}

\appendix

\end{document}